\documentclass[aps,prl,twocolumn,showpacs,preprintnumbers,amsmath,amssymb,superscriptaddress]{revtex4}

\usepackage{amsmath,amsthm,amsfonts,amscd} 
\usepackage{eucal}
\usepackage{graphicx}
\usepackage{braket}			% Allows use of Dirac notation
\usepackage{color}		% Allows \textcolor{red}{Hello world}
\usepackage{pxfonts}	% Needed for \medcirc and some other symbols
\usepackage{MnSymbol}	% Needed for \smalltriangleup and others
\usepackage{ulem} 		% Proofing tools such as cross-thorough \sout{blah blah}
\usepackage{upgreek}	% For non-italicized lower-case greek letters, like $\upmu$

\begin{document}

\title{Coalescence of Single Photons Emitted by Disparate Single Photon Sources: The Example
of InAs Quantum Dots and Parametric Down-Conversion Sources}
%\maketitle

\author{Sergey V. Polyakov}
\affiliation{Joint Quantum Institute, National Institute of Standards and Technology 
				\& University of Maryland, Gaithersburg, MD, USA.}
\author{Andreas Muller}
\altaffiliation[AM is currently in ]{the Department of Physics, University of South Florida, Tampa FL.}
\affiliation{Joint Quantum Institute, National Institute of Standards and Technology 
				\& University of Maryland, Gaithersburg, MD, USA.}

\author{Edward B. Flagg}%
\affiliation{Joint Quantum Institute, National Institute of Standards and Technology 
				\& University of Maryland, Gaithersburg, MD, USA.}
				
				\author{Alex Ling}%
\altaffiliation[AL is currently at ]{the Centre for Quantum Technologies, Singapore 117543.}
\affiliation{Joint Quantum Institute, National Institute of Standards and Technology 
				\& University of Maryland, Gaithersburg, MD, USA.}

\author{Natalia Borjemscaia}%
\affiliation{Joint Quantum Institute, National Institute of Standards and Technology 
				\& University of Maryland, Gaithersburg, MD, USA.}
\affiliation{Physics Department, Georgetown University, Washington, DC, USA}

\author{Edward Van Keuren}
\affiliation{Physics Department, Georgetown University, Washington, DC, USA}
\author{Alan Migdall}%
\affiliation{Joint Quantum Institute, National Institute of Standards and Technology 
				\& University of Maryland, Gaithersburg, MD, USA.}
\author{Glenn S. Solomon}
\email[Correspondence: ]{glenn.solomon@nist.gov}
	\affiliation{Joint Quantum Institute, National Institute of Standards and Technology 
				\& University of Maryland, Gaithersburg, MD, USA.}

\date{\today}

\begin{abstract}
Single photons produced by fundamentally dissimilar physical processes will in general not be indistinguishable. We show how photons produced from a quantum dot and by parametric down-conversion in a nonlinear crystal can be manipulated to be indistinguishable.  The measured two-photon coalescence probability is 16\%, and is limited by quantum-dot decoherence. Temporal filtering to the quantum dot coherence time and accounting for detector time response increases this to 61\% while retaining $25\%$ of the events. This technique can connect different elements in a scalable quantum network.
%Recent progress in quantum information has added practical interest to this fundamental problem.
\end{abstract}
% ~130 characters per line, 600 characters max (including whitespace).

	% 78.67.Hc Quantum Dots
	% 42.50.Ar Photon statistics and coherence theory 
	
%\keywords{}

\maketitle

%In quantum mechanics, particles in identical states are indistinguishable, giving rise to effects with no classical analog \cite{new8}. 	
%The indistinguishability of particles in identical states leads to phenomena unique to quantum mechanics \cite{new8}. Recent progress in quantum information has added practical interest to this fundamental problem. 
%Establishing quantum networks requires interconnects between distinct components.  It is unlikely that these different components will be composed of the same physical system. 

In quantum mechanics, particles in identical states are indistinguishable, giving rise to effects with no classical analog \cite{new8}.
For instance, the bosonic nature of light insures that upon interference two indistinguishable photons will coalesce into a single inseparable state.  
Using this coalescence property, a high degree of indistinguishability was demonstrated from a single source of correlated photons [2,3].  
Photons from a single quantum system, for instance an atom or ion, a molecule or a single semiconductor quantum dot (QD) have been shown to be indistinguishable \cite{1a,2a,3a}.  
In addition, photons from separate, but nominally identical sources, such as two ions, two atoms, or two QDs can produce indistinguishable states \cite{4,5,6,7}.  
%Photons from separate, but similar sources, for instance, as in recent experiments with two ions or two atoms, can also coalesce with high fidelity [4,5], while photons from less similar sources, such as semiconductor quantum dots (QDs) do so with less fidelity [6,7]. 
Highly dissimilar photon sources will not produce identical photons except by random coincidence.  However, it is in principle possible to manipulate photon characteristics without loss of quantum coherence. 

Recent progress in quantum information has created additional interest in this fundamental problem.
Establishing quantum networks requires interconnects between distinct components and it is unlikely that these components will be composed of the same physical system.  For instance, quantum gates require good coupling \cite{9}, while quantum memories \cite{10} call for long-lived, decoherence-free (minimal coupling) materials; these processing and storage elements are best implemented by matter states \cite{11,12,13,14,15}. 
Connecting such elements using discrete, single photons is one promising integration option.  Here, single photon sources are necessary -- coherent sources will not in general be adequate, and it will be important that the single photons be indistinguishable. Yet, single photons produced by fundamentally dissimilar physical processes will in general not be indistinguishable.  
%Here we show for the first time how single photons produced from fundamentally dissimilar physical processes can be manipulated to be indistinguishable.

%Different quantum systems have different attributes that can be leveraged for quantum information. 
%For instance, quantum gates require strong coupling \cite{9}, while quantum memories \cite{10} call for long-lived, decoherence-free materials; these processing and storage elements are best implemented by matter states \cite{11,12,13,14,15}. 
%These QD and nonlinear crystal sources are promising candidates for matter and photon elements in quantum information technologies, such a quantum network, quantum repeater, and linear-optics quantum computing.  
Here we show that photons from two dissimilar single-photon sources can be manipulated so that they coalescence upon interference.
The two single photon sources used here are a solid-state QD decay and parametric down-conversion (PDC) generation.
Semiconductor QDs are promising nodes for qubit processing because of their large dipole moments, compatibility with well-developed semiconductor processing and scalability. PDC sources are promising quantum communication interconnects.  They have high fidelity, can propagate quickly, and have low coupling to the environment \cite{16, 17, 18}.
%We show how a  QD exciton decay and parametric down-conversion (PDC) generation can be manipulated so that they become indistinguishable. 
%We use both heralded and unheralded PDC sources, and show the coalescence probability is 16\%, limited by decoherence in the QD.   
%Photons produced from different QDs can be made indistinguishable \cite{6,7}, albeit with reduced indistinguishability due to decoherence. 
%This decoherence can be reduced by employing higher Purcell factor cavities, but leads to additional challenges such as spectral alignment of multiple cavities with multiple QD transitions. 
%In contrast, entangled photons created in PDC have excellent coherence properties. 
Unlike QD photons, PDC photons are not antibunched \cite{19}, but because photons are produced in pairs, when operated at low pump power the detection of one photon in one channel heralds one and only one photon in the remaining channel.  This produces a near perfect heralded single-photon PDC source.
The photons from QD and PDC differ in their spectral and
temporal properties by several orders of magnitude and thus the natural photonic wavepacket
overlap will be minimal. 
%However, we can manipulate these wavepackets to match each other and we demonstrate this here.

To quantitatively assess the photon indistinguishability, we combine them on an optical beam splitter and measure the second-order cross-correlation of the outputs [2],
as shown in Fig. 1(a). To ensure temporal synchronization, the single-photon sources are excited by the same 76 MHz pulsed
laser (at 820 nm with 8 ps pulse duration).
A parametric frequency doubler generates the pump for the PDC process. 
A periodically-poled KTP crystal generates photon pairs \cite{22}. Detection of a photon
at 740 nm heralds the presence of a photon at 918 nm. The two PDC photons are separated on a
polarizing beam splitter and coupled into single-mode fibers. 
The spectral linewidth of the 918 nm PDC
photon is 1.5 nm  ($\Delta \nu  \approx$ 533 GHz) and is more than 500 times broader than the 918 nm QD photon 
($\Delta \nu \approx 1$ GHz).
To match the spectral linewidths of a PDC photon at 918 nm while maintaining high heralding efficiency, we use a tunable spectral
filter with active stabilization, as well as a separate filter on the heralding photon.

\begin{figure}[h!]
\includegraphics[width=3.0in]{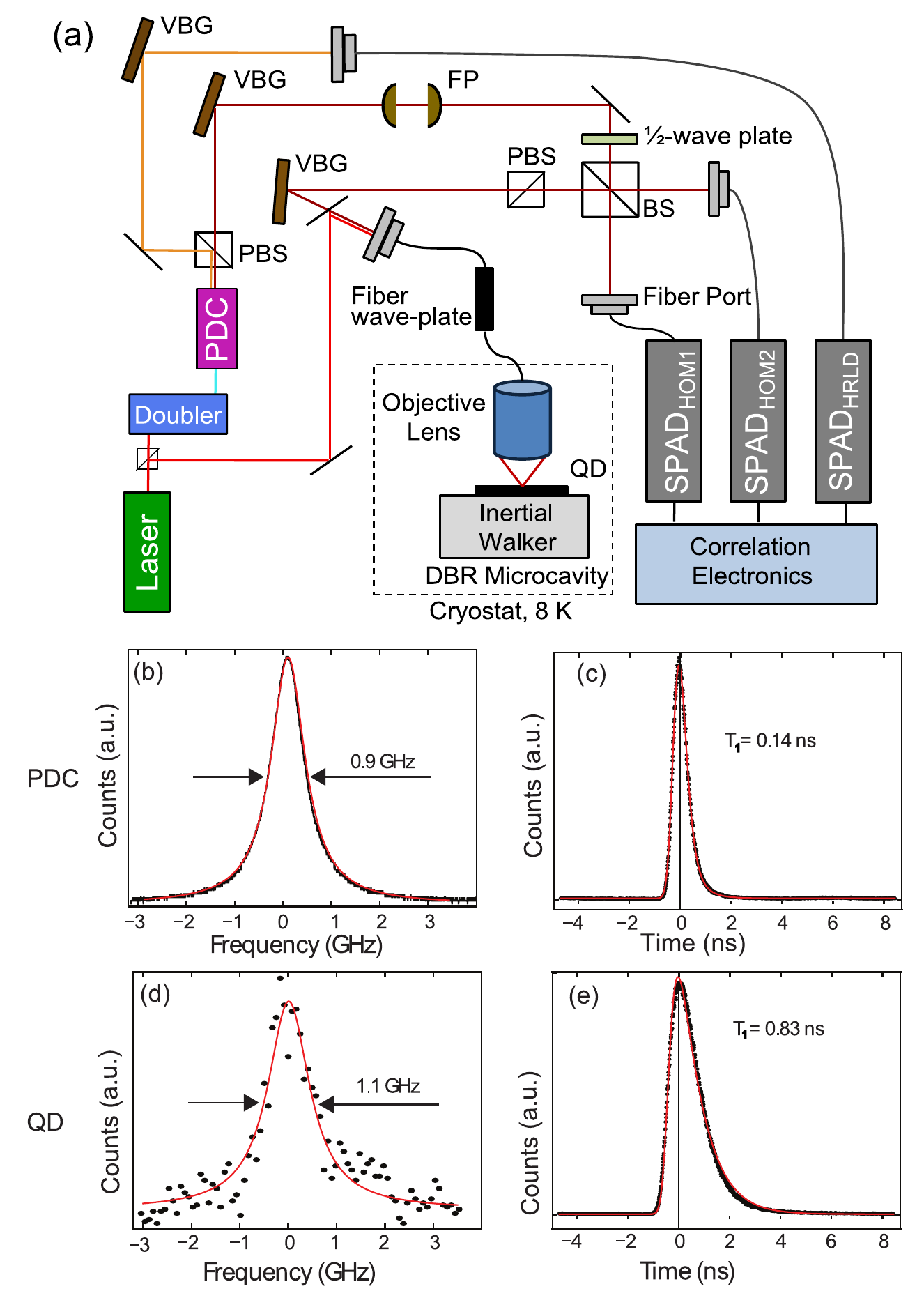}
\caption{\label{fig1} (color online) (a) Experimental apparatus. VBG: volume-Bragg grating; FB: Fabry-Perot cavity; PBS: polarizing beam splitter.  Characterization of sources: (b) linewidth and (c) temporal properties of PDC; (d) linewidth and (e) temporal properties of QD. Black dots: measured values; red thin lines: Lorenztian (b,d) or exponential decay fits including the detector response time (c,e).}
\end{figure}

The spectral properties of the PDC photon are set by the transmittance properties of our spectral filter.
The Fabry-Perot cavity produces a comb of transmission peaks with FWHM of $\Delta \nu_{\rm{PDC}}$= ($\rm0.9\pm 0.1$) GHz (Fig.
1(b)), closely matching that of the QD. We also confirm that the resulting temporal pulse duration is 0.14
ns (Fig. 1(c)) by making a direct measurement with our detector and accounting for the detector's time
resolution.

The strain-induced InAs QD is embedded in a planar distributed Bragg reflector microcavity (DBR) of alternating GaAs and AlAs layers.
The spectral character of the QD is shown in Fig. 1(d). 
It has a linewidth of 1.1 GHz, implying a coherence time of $T_2$ = 0.29 ns. 
Time-dependent fluorescence of the QD is shown in Fig. 1(e),
which when fit by an exponential decay and a finite detector time resolution, yields a lifetime of $T_1$ = 0.83
ns. Since
$T_2 < 2T_1$, the coherence
time is not lifetime limited. We confirm that this QD transition consists of two fine-structure split
lines that have orthogonal polarization, and thus the photon is the result of spontaneous
emission decay from the ground-state optically active neutral exciton. For indistinguishability measurements, one emission line is selected
by spectral and polarization filtering. The experimental setup with the PDC input blocked can be used
for an auto-correlation measurement \cite{20} of the QD photon. The second-order autocorrelation of QD photons is shown in Fig. 2. 
We observe that the counts are largely suppressed for the
time delay of 0, demonstrating the single-photon character of our source. The number of counts in the
central, zero-delay peak integrated over the peak duration is 16.5\% of the averaged integrated counts
for adjacent peaks. No background subtraction has been applied to the data. This is well below the 50\% that can only be achieved by
a single-photon source \cite{21}. 
The PDC source is even better, with the integrated zero-peak $g^{(2)}(\tau)$ estimated to be $< 10^{-3}$.

\begin{figure}[t!]
\includegraphics[width=3.1in]{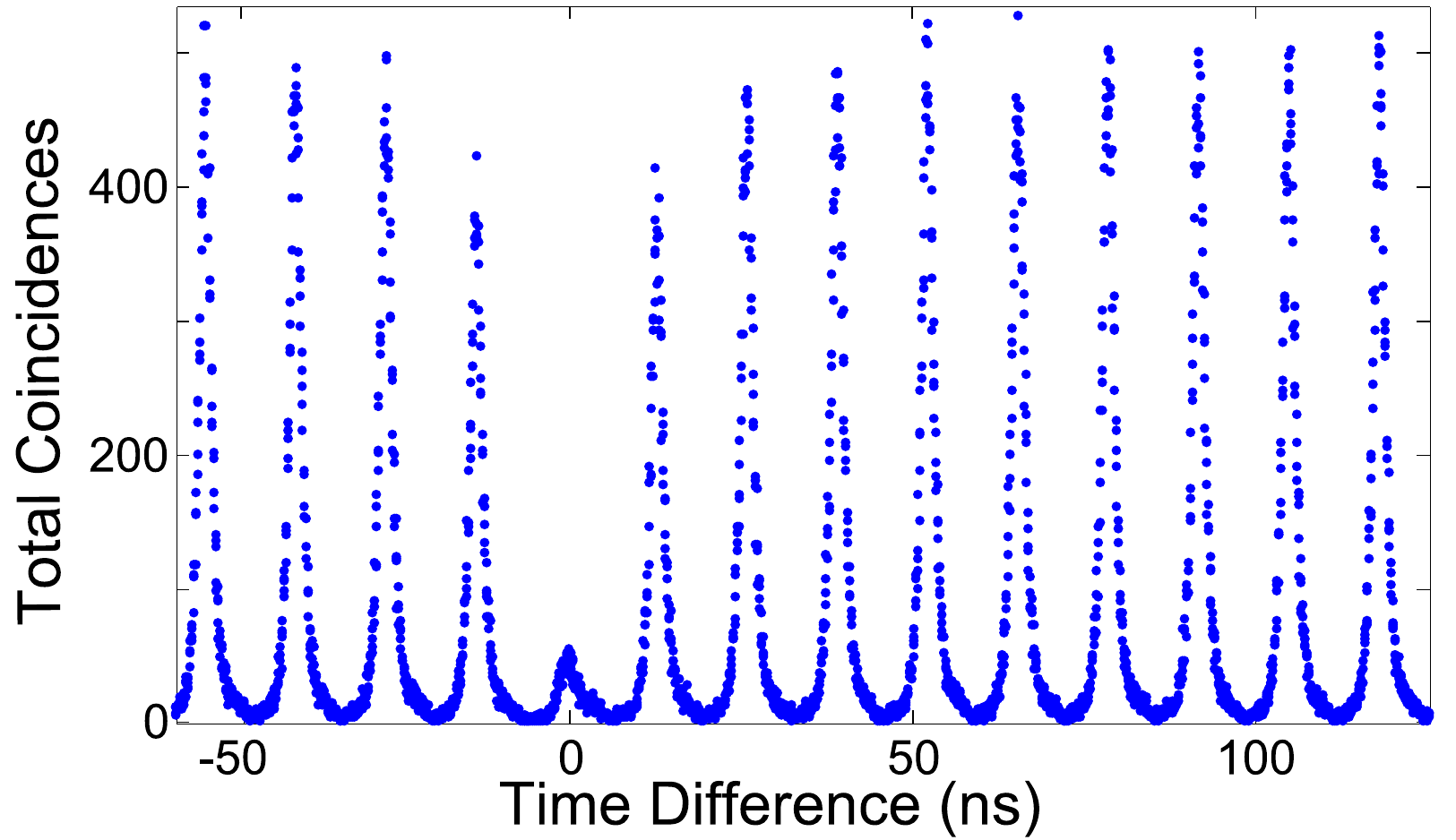}
\caption{\label{fig2} Measured second-order autocorrelation of the QD.}
\end{figure}

The two single-photon states are then sent to interfere at a 50/50 nonpolarizing
beamsplitter. The detection of the heralding photon triggers a measurement of the PDC - QD Hong-Ou-Mandel (HOM)
interference \cite{2}. The beam paths are adjusted to ensure spatial and
temporal overlap of the single-photon states. Two single-photon avalanche detectors (SPADs) monitor
the outputs of the beamsplitter. Another SPAD monitors the heralding output of the PDC crystal. A half-wave
plate placed in the PDC arm can rotate the PDC photon's polarization and switch the interference on (parallel polarization) and
off (orthogonal).  Timestamps of detections relative to the laser pulse for all SPADs are recorded for
statistical analysis.

To characterize the indistinguishability of the QD and PDC photons, we measure the second-order
intensity cross-correlation of the output ports of the interferometer conditioned upon a heralding
detection. We restrict our measurement to the trials in which a detection in the heralding channel has
occurred ({\it heralded trials}), since this additional condition yields highly antibunched  PDC photons with increased emission probability \cite{24a}. 
Fig. 3(a)
illustrates a conditional cross-correlation measurement based on this heralded detection. Non-heralded
trials are discarded and the remaining trials are renumbered. The event trial difference is $\Delta n = m - l$,
where {\it  l,m} are the new trial numbers corresponding to detections in different HOM channels.  
A measured detection occurs at time $\Delta t_{i,j}$ on a detector {\it i} (1 or 2) within heralded trial {\it j}. 
The time difference within $\Delta n$ between two heralded detections then becomes $\tau = \Delta t_{2,m} - \Delta t_{1,l}$.
If the detections
occur during the same trial ($\Delta n$ = 0) interference effects are possible. 
For all other $\Delta n$, no interference effects occur. 
For example in Fig. 3(a), one event with $\Delta n =$ 0 occurs in heralded trial 3. Also, one event
with $\Delta n = 1$ and one event with $\Delta n = 2$ occur due to the detection at 
HOM 1 in heralded trial 1, and detections at HOM 2 in heralded trials 2 and 3.
Under these conditions the QD photon count rate is 0.5 of the heralded PDC rate, and this ratio is uniform throughout all heralded trials.
In Fig. 3(b,c) we construct the conditional
second-order cross-correlation function for the orthogonal and parallel polarizations of the input single-photon
states, C$_{\bot,\|}$ Fig. 3(b,c). Note that no background correction was applied. 
The area of the $\Delta n=0$ peak of the second order cross-correlation function (Fig. 3(b)) is smaller than half that of the
neighboring peaks for both polarizations, because the inputs are single-photon states.  
For parallel polarizations of the inputs the height of the $\Delta n=0$ peak is lower than for perpendicular
polarizations, indicating two-photon coalescence. We emphasize that this coalescence occurs
with two independent single-photon sources of drastically different physical nature. Therefore, the
interfering single-photon states share no common or even comparable history.

\begin{figure}[h!]
\includegraphics[width=2.85in]{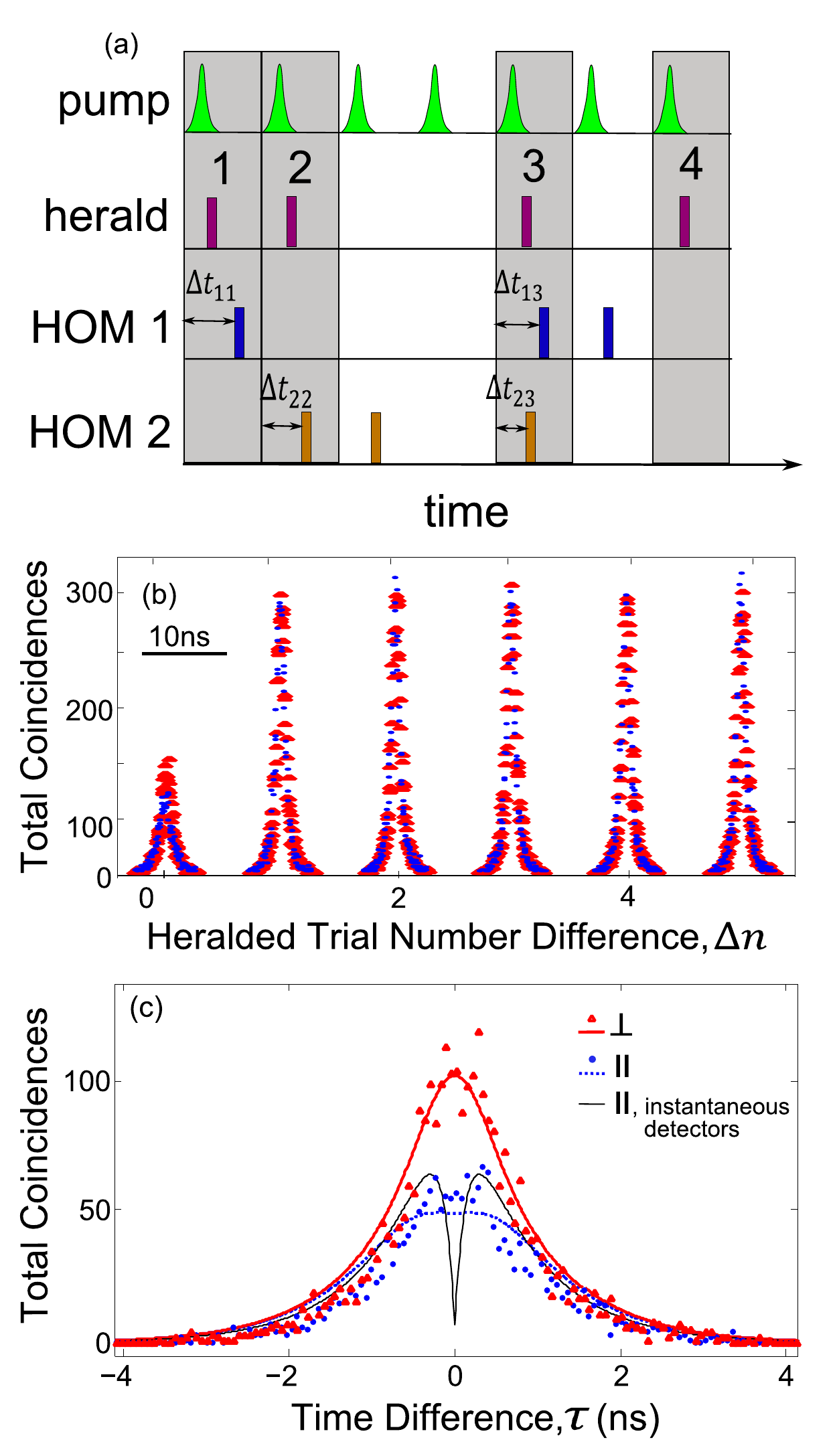}
\caption{\label{fig3} (color online) Conditional second-order cross-correlation with a heralded PDC source. (a) To compute conditional second-order cross-correlation, trials with no heralding detections are first discarded. (b) Measured conditional second-order cross-correlation. (c) Close-up of $\Delta n = 0$ peak.  
Experimental data: red triangles: perpendicular polarization; blue circles: parallel polarization. Modeling of the data: red solid curve: perpendicular polarization; blue dotted curve: parallel polarization with the realistic time response of detectors; black thin curve: parallel polarization in the limit of instantaneous detectors.}
\end{figure}

Fig. 3(c) shows a close-up of the $\Delta n=0$ peak.  
For perpendicular polarization (red triangles) the two single-photon sources produce fully distinguishable photons. 
For parallel polarization (blue circles), two-photon interference suppresses the peak.
The coalescence is most pronounced in the center of the peak and disappears towards the tails. 
%\sout{This reduction likely occurs because the QD coherence time is significantly shorter than the lifetime ($T_{2}<$ 2$T_{1}$). }
% \sout{Thus, photons emitted by this QD begin to lose their ability to interfere for time delays that are on the order of $T_{2}$. }
%\sout{The theoretical description of this effect \cite{3a} prescribes that at short delays, $\tau<< T_{2}$, the interference should be nearly perfect, producing a deep, narrow dip} \cite{6,7,23}. 
%\sout{However, because the resolution of the detectors used for this experiment is also on the order of $T_{2}$, such a dip cannot be resolved fully.}

% \sout{We use the model of Ref.\cite{3a}
%with the measured parameters of our QD and PDC photons to determine if the imperfect coalescence seen in Fig. 3 is caused by the coherence time of the QD not being lifetime limited \cite{25}. 
%The model also takes into account the non-zero probability of the QD to produce more than one photon.
%We stress that our model assumes no other mechanisms that could limit the HOM visibility. 
%Curves produced by the model are shown in Fig. 3(c) for different polarizations and for real or ideal detectors.
%The model adequately describes the effect of the coalescence decrease with delay $\tau$. }
To quantitatively determine the degree of indistinguishability, we define the probability of coalescence of
the photons from the two sources as: $P_{c} = \left(A_{\bot} - A_{\|}\right)/A_{\bot}$, 
where $A_{\bot,\|}= \int C_{\bot,\|} \left( 0,\tau \right)d\tau$ \cite{6} is the number of
counts in the $\Delta n =0$ peak of the second-order cross-correlation function integrated over the full
temporal extent of the peak. From the experimental data presented in Fig. 3 we get $P_{c}=(16\pm3)\%$ \cite{k}.
We use the model of Ref. \cite{3a}
with the measured parameters of our QD and PDC photons to determine a theoretical maximum for coalescence \cite{25}. 
The model also
takes into account the non-zero probability of the QD to produce more than one photon.
We stress that
our model assumes no other mechanisms that could limit the HOM visibility. 
From the measured source parameters, the model predicts a theoretical maximum for coalescence, $P_{c,max}=(27\pm3)\%$.
Curves produced by the model are shown in Fig. 3(c) for different polarizations and for real or ideal detectors.
The model adequately describes the effect of the coalescence decrease with delay $\tau$. 
The decrease occurs because the
QD coherence time is significantly shorter than the lifetime ($T_{2}<$ 2$T_{1}$). 
Thus, photons emitted by this QD begin to lose their ability to interfere for time delays that are on the
order of $T_{2}$.
The theoretical description of this effect \cite{3a} prescribes that at short delays, $\tau<< T_{2}$, the interference should be nearly perfect, producing a deep, narrow dip \cite{6,7,23}. 
However, because the resolution of the detectors used for this
experiment is also on the order of $T_{2}$, such a dip cannot be resolved fully.
The value for $P_{c}$ is similar to that of Ref. \cite{6} even though the PDC is decoherence free.  
While we match the linewidths ($\propto T^{-1}_{2}$) of the two sources, the $T_{1}$ times are not similar because for the QD photon $T^{-1}_{2}\neq (2T_{1})^{-1}$.  
%\sout{In Ref. \cite{6} the sources temporally overlap up to the dephasing limit, while here they} 
Thus, the sources overlap up to the smaller $T_{1}$ -- that of the PDC.  
However, the $P_{c}$ reported here is larger than would be obtained with two similar QDs because here the PDC source is decoherence free.

\begin{figure}[h!]
\includegraphics[width=2.75in]{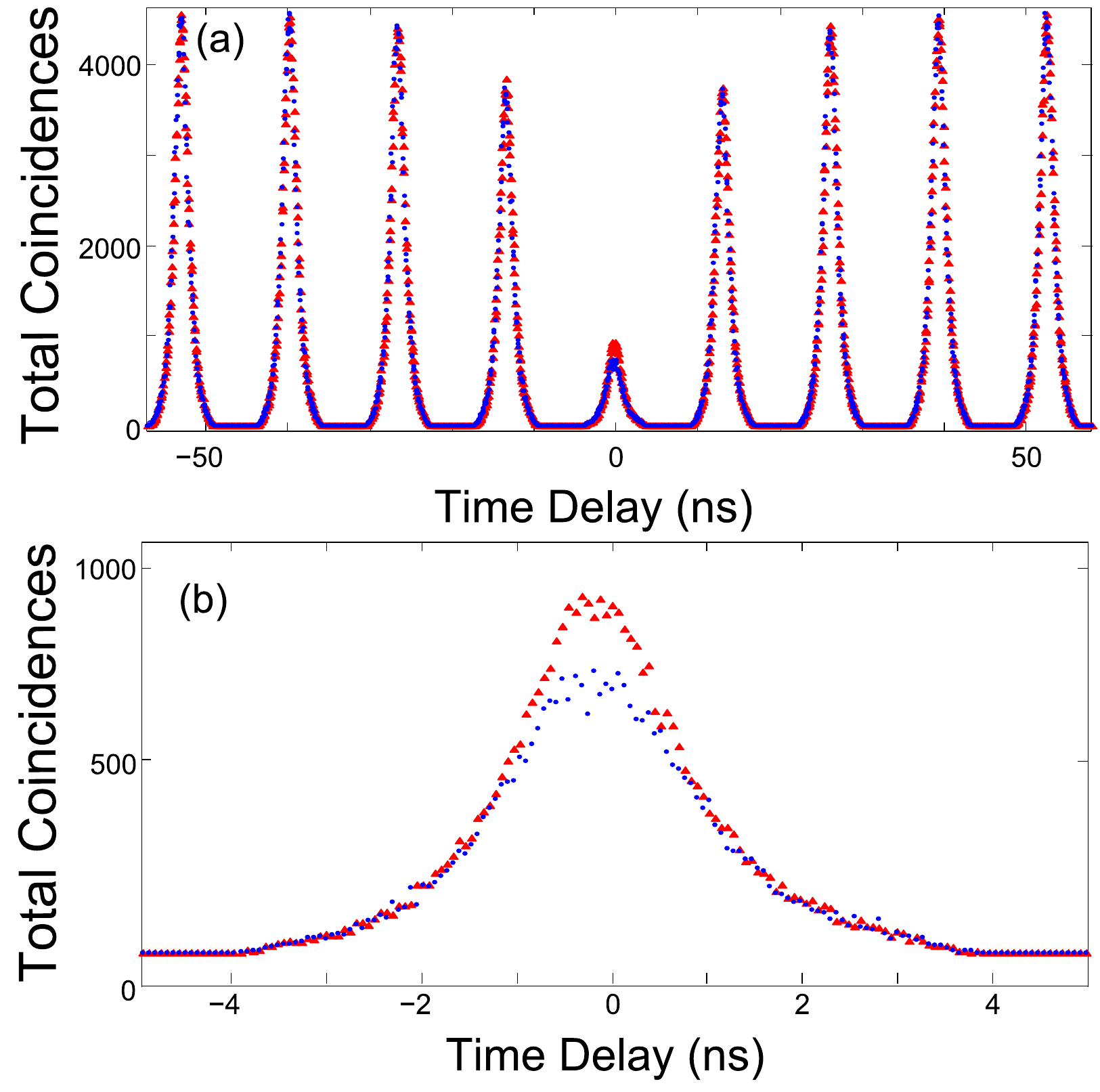}
\caption{\label{fig4} (color online) Measured second-order cross-correlation function with an unheralded PDC source. (a) Same as Fig. 3 (b), however the PDC source is not heralded. The shape of the cross-correlation function is mainly determined by an autocorrelation of the strongest emitter, the QD ({\it c.f.} Fig. 2). (b) A close-up on the peak at zero time delay. The two-photon coalescence probability is reduced to $13 \%$, {\it c.f.} Fig. 3 (c).}
\end{figure}

The model suggests that the best coalescence is achieved by postselecting for short delay times: $\tau \rightarrow 0$.
We define a post-selective coalescence probability, $P_{c}(0) = \left[ C_{\bot}(0,0) - C_{\|} (0,0) \right]$ $/ C_{\bot}(0,0)$ which we measure to be $P_{c}(0)=(42\pm5)\%$ because of the time jitter of the detectors. 
For infinitely fast detectors we would have measured $P_{c}(0)=86\%$, while the model predicts $P_{c,max}(0)=97$\%. 
The main reason this differs from unity is
that our QD is not a perfect single-photon source. 
Thus with appropriately fast detectors we could increase the coalescence by discarding
detection events after the measurement (post selection) for $\tau$ greater than some threshold, $\tau_{0}$. 
However, doing so for $\tau_{0} \rightarrow 0$
would be of little physical significance, because for truly instantaneous photodetections, single photons
will always coalesce \cite{26}. 
Instead, we can take advantage of the initial period after excitation where the temporal overlap is large and the QD photon remains coherent by temporally gating either the sources or the detectors around the laser pump pulse.
%\sout{\textcolor{red}{The timestamping of the data relative to the laser pulse} \textcolor{red}{ allows us to simulate such gating with the current data.}
%\textcolor{red}{Gating with a window equal to the PDC photon $T_{1}$}
%\sout{Doing so with a 140 ps window, the PDC photon T$_{1}$,}
%temporal gating either on the sources or the detectors around the laser pump pulse.}
When accounting for our detector resolution and adding temporal filtering in the experiment with a 290 ps window - the QD coherence time, we obtain an experimental coalescence probability of $P_{c, fraction} = 61 \%$ while retaining $25\%$ of the events.  
Reducing the filtering window to 140 ps - the temporal extent of the PDC photon, increases $P_{c, fraction}$ to $75 \%$, while now retaining $10\%$ of the events.  

If the PDC and QD photons were truly
indistinguishable, a click on either (but not both) of the HOM detectors would erase ``which path"
information and create potentially useful entanglement between the second PDC field and the state of
the QD \cite{15,27}. Thus, it is important to assess the degree of indistinguishability between the
two fields in the unheralded case. In Fig. 4 we see that the cross-correlation function in this case closely
follows that of the heralded case. 
The major difference from the heralded case is that
the shape of the cross-correlation function, apart from the interference at the zero peak, is mainly
determined by the autocorrelation of the QD; note the decrease in the near-zero peaks in both Fig. 2 and Fig. 4(a).  
%Fig. 2, and note the similar behavior of the peaks immediately adjacent to the zero peak). 
This observation is not surprising, because the probability to
detect a single photon emitted by a QD is much higher ( 30,000 s$^{-1}$) than that produced by an unheralded PDC (300 s$^{-1}$). The
coalescence figures in the unheralded case are $P_{c}=(13.2\pm0.8)\%$ and $P_{c}(0)=25\%$; they are similar but lower than those of the heralded PDC field case.

In conclusion, we have demonstrated a sizable coalescence between photons from two sources of a
different nature by manipulating the individual fields. We have also shown that the main reason for
reduced coalescence is decoherence of the QD used in this experiment. 
Importantly, this coalescence can be improved with gating around the laser pump pulse or reduced QD dephasing. 
Reduced dephasing has been demonstrated in separate studies \cite{1a}
in which similar QDs experienced
large Purcell factors due to tighter optical-mode confinement. 
%In particular, it has recently been shown
%that such cavities can be engineered around this type of QDs, guaranteeing QD transition and cavity
%alignment \cite{28}. 
In concert with such a device, the technique presented here can be used to produce indistinguishable
photons from different elements in a quantum network.

We acknowledge partial support from the NSF Physics Frontier Center at the Joint Quantum Institute, and thank Elizabeth Goldschmidt for helpful discussions.\\


\begin{thebibliography}{99}
\bibitem{k} All uncertainties stated are one standard deviation.
\end{thebibliography}
\end{document}